\documentclass{desyproc}

\begin{document}
\title{Recent Results in Polarized Proton-Proton Elastic
Scattering at STAR.
}

\author{{\slshape Leszek Adamczyk$^1$}\\[1ex]
$^1$Faculty of Physics and Applied Computer Science, AGH-UST Cracow, Poland}

\contribID{8}


\acronym{EDS'09} 

\maketitle

\begin{abstract}
RHIC is the only spin-polarized proton collider ever built. With a special optics run of $\beta^\star \approx 22$ m STAR detector system is suitable for an investigation of the spin dependence of elastic proton-proton scattering.  This is a brief summary of measurements of spin asymmetries at the center of mass energy $\sqrt{s}=200$ GeV 
and in the four-momentum transfer squared $-t$ range $0.003 < -t < 0.035$ GeV$^2$ by the STAR experiment at RHIC. 
\end{abstract}

\section{Formalism of proton-proton elastic scattering }
Elastic proton-proton scattering is described by five independent helicity amplitudes. 
Two helicity non-flip ones $\phi_1(s,t) \propto \langle ++ | M | ++ \rangle$ and
$\phi_3(s,t) \propto \langle +- | M | +- \rangle$,  two double helicity-flip ones 
$\phi_2(s,t) \propto \langle ++ | M | -- \rangle$  and
$\phi_4(s,t) \propto \langle +- | M | -+ \rangle$ and one single helicity-flip amplitude
$\phi_5(s,t) \propto \langle ++ | M | +- \rangle$ (see Ref. ~\cite{HelAmpl}). Each amplitude consists of hadronic and electromagnetic parts $\phi_i = \phi_i^{em} + \phi_i^{had}$.
The azimuthal angle dependence of the cross section of vertically polarized protons are given~\cite{Spin} by:
$$2\pi\frac{d^2\sigma^{\updownarrow \updownarrow}}{dt d\phi} = \frac{d\sigma}{dt} \left(1+A_N(t)(P^{\updownarrow}_Y+P^{\updownarrow}_B) \cos(\phi) + P^{\updownarrow}_Y P^{\updownarrow}_B[A_{NN}(t)\cos^2(\phi) +  A_{SS}(t) \sin^2(\phi)] \right)$$ 
where $d\sigma/d t$ is spin average cross section, "$\updownarrow$" (either "$\uparrow$" or "$\downarrow$") indicate the spin direction of transversely polarized colliding proton beam bunches,  $P^{\uparrow}_Y(P^{\downarrow}_Y)$ and $P^{\uparrow}_B(P^{\downarrow}_B)$ are the beam polarizations for the two colliding beams (called Blue and Yellow), $A_N, A_{NN}$ and 
$A_{SS}$ are spin asymmetries.    The spin asymmetries relate to the helicity amplitudes as follows~\cite{HelAmpl}
$$A_N(s,t)\frac{d\sigma}{dt} = \frac{-4\pi}{s^2} \text{Im}\left(\phi_5^\star(\phi_1+\phi_2+\phi_3-\phi_4)\right)$$
$$A_{NN}(s,t)\frac{d\sigma}{dt} = \frac{4\pi}{s^2} \left(2|\phi_5|^2 + \text{Re}(\phi_1^\star\phi_2-\phi_3^\star\phi_4)\right)$$
$$A_{SS}(s,t)\frac{d\sigma}{dt} = \frac{4\pi}{s^2} \text{Re} \left(\phi_1\phi_2^\star+\phi_3\phi_4^\star\right)$$

For $|P^{\downarrow}_Y| =|P^{\uparrow}_Y| = P_Y$ and $|P^{\downarrow}_B| =|P^{\uparrow}_B| = P_B$ some combination of measurements allows extraction of spin asymmetries from uncorrected for inefficiencies event numbers ($N$) normalized by respective luminosities($L$).
$$ \frac{  A_{N} (P_Y+P_B)\cos(\phi)}{1+\delta(\phi)}  =  \frac{N^{\uparrow\uparrow}/L^{\uparrow\uparrow} - N^{\downarrow\downarrow}/
L^{\downarrow\downarrow}}{N^{\uparrow\uparrow}/L^{\uparrow\uparrow} + N^{\downarrow\downarrow}/L^{\downarrow\downarrow} }
; \ \ \frac{A_{N} (P_B-P_Y)\cos(\phi)}{1-\delta(\phi)}  =  \frac{N^{\uparrow\downarrow}/L^{\uparrow\downarrow} - N^{\downarrow\uparrow}/
L^{\downarrow\uparrow}}{N^{\uparrow\downarrow}/L^{\uparrow\downarrow} + N^{\downarrow\uparrow}/L^{\downarrow\uparrow} }
$$
$$ \delta(\phi) = P_Y P_B[A_{NN}\cos^2(\phi) + A_{SS} \sin^2(\phi)]  = \frac{ (N^{\uparrow\uparrow}/L^{\uparrow\uparrow} + N^{\downarrow\downarrow}/L^{\downarrow\downarrow}) - (N^{\uparrow\downarrow}/L^{\uparrow\downarrow} + N^{\downarrow\uparrow}/L^{\downarrow\uparrow} )}
{ ( N^{\uparrow\uparrow}/L^{\uparrow\uparrow} + N^{\downarrow\downarrow}/L^{\downarrow\downarrow}) + (N^{\uparrow\downarrow}/L^{\uparrow\downarrow} + N^{\downarrow\uparrow}/L^{\downarrow\uparrow} )}
$$

Double spin symmetries were measured based on fits of the formula:  
$$  \delta(\phi) = P_Y P_B\left[ \frac{A_{NN}+A_{SS}}{2}+ \frac{A_{NN}-A_{SS}}{2} \cos(2\phi) \right]$$
to the azimuthal angle distributions.

Single spin asymmetry was calculated using geometric means~\cite{Geom} so-called square root formula for each pair of $\phi$ and $\pi-\phi$ bins in the range $\pi/2<\phi<\pi/2$
$$ 
  \frac{A_{N} (P_Y+P_B)\cos(\phi)}{1+\delta(\phi)}  =  
 \frac{\sqrt{N^{\uparrow\uparrow}(\phi)N^{\downarrow\downarrow}(\pi-\phi)} -
              \sqrt{N^{\downarrow\downarrow}(\phi)N^{\uparrow\uparrow}(\pi-\phi)}}
             {\sqrt{N^{\uparrow\uparrow}(\phi)N^{\downarrow\downarrow}(\pi-\phi)} +
              \sqrt{N^{\downarrow\downarrow}(\phi)N^{\uparrow\uparrow}(\pi-\phi)}} 
        $$
In the square root formula, the relative luminosities of different spin direction cancel out.
\section{Experiment}
\begin{wrapfigure}{r}{0.4\textwidth}
  \vspace{-20pt}
  \begin{center}
    \includegraphics[width=0.38\textwidth]{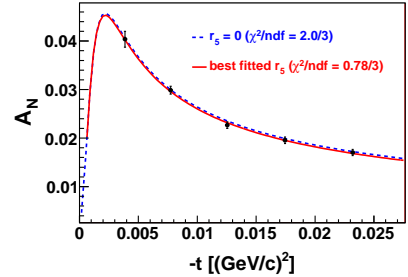}
  \end{center}
  \vspace{-20pt}
  \caption{Single spin asymmetry $A_N$ for five $−t$ intervals. Vertical error bars show statistical uncertainties. The dashed curve corresponds to theoretical calculations without hadronic spin-flip and the solid one represents the $r_5$ fit} \label{AN}
  \vspace{-10pt}
\end{wrapfigure}

The protons, which scatter elastically at small angles ($< 2$ mrad), follow the optics of the RHIC magnets and are
detected by a system of detectors placed close to the beam inside movable vessels known as “Roman Pots” (RPs) \cite{RP}.
The Roman Pot stations are located on either side of the STAR interaction point (IP) at 55.5 m and 58.5 m with
horizontal and vertical insertions of the detectors, respectively. 
The data were taken during four dedicated RHIC stores in 2009 with special beam optics
of $\beta^\star = 22$  m in order to minimize the angular divergence at the IP \cite{beam}.
The selection of elastic events was based on the collinearity of the scattered proton tracks.
\section{Single spin asymmetry}
The contribution of two double spin-flip amplitudes($\phi_2$ and $\phi_4$) to the $A_N$ asymmetry is small. Thus the main contribution to $A_N$ is given by:
 $$A_N(s,t)\frac{d\sigma}{dt} \approx \frac{-4\pi}{s^2} \text{Im}\left(\phi_5^{em \star}(\phi_1^{had}+\phi_3^{had}) + \phi_5^{had \star}(\phi_1^{em}+\phi_3^{em})\right)
\label{eq}
$$

The electromagnetic amplitudes are fully calculable in QED, leading non helicity-flip hadronic contribution($\phi_1^{had}$ and $\phi_3^{had}$) to the $A_N$ can 
be constrain by total proton-proton cross section.
Thus there is precise prediction on $A_N$ for $\phi_5^{had}=0$. Any deviation from above prediction indicates contribution  from hadronic spin-flip processes caused by
Reggeon or Pomeron.
The measured values of $A_N$ are presented in Fig. \ref{AN} together with parameterizations based
on theoretical calculations \cite{HelAmpl}: the dashed line corresponds to no hadronic spin-flip contribution, while the solid line is the result of the fit using, 
$r_5$, the measure of the ratio of the hadronic single spin-flip amplitude to hadronic single
non-flip amplitudes as a free parameter. The obtained
values Re $r_5 = 0.0017\pm 0.0063$ and Im $r_5 = 0.007\pm 0.057$ are consistent with the hypothesis of no hadronic spin-flip
contribution at the energy of this experiment. The high accuracy of the current measurement provides
strong limits on the size of any hadronic spin-flip amplitude at this high energy.

\section{Double spin asymmetries}
The contributions of single spin-flip amplitude($\phi_5$) to the $A_{NN}$ and $A_{SS}$ asymmetries are small. Thus the main contribution to double spin asymmetries 
are given by:
$$\frac{A_{NN}+A_{SS}}{2}  \frac{d\sigma}{dt} \approx \frac{4\pi}{s^2} \text{Re}(\phi_1\phi_2^\star)$$
$$\frac{A_{NN}-A_{SS}}{2}  \frac{d\sigma}{dt} \approx -\frac{4\pi}{s^2} \text{Re}(\phi_3\phi_4^\star)$$ 
$A_{NN}$ and $A_{SS}$ are sensitive to the Odderon contribution \cite{Oderon},
the hypothetical counterpart of the Pomeron that carries odd charge parity.

\begin{wrapfigure}{r}{0.5\textwidth}
  \vspace{-20pt}
  \begin{center}
    \includegraphics[width=0.48\textwidth]{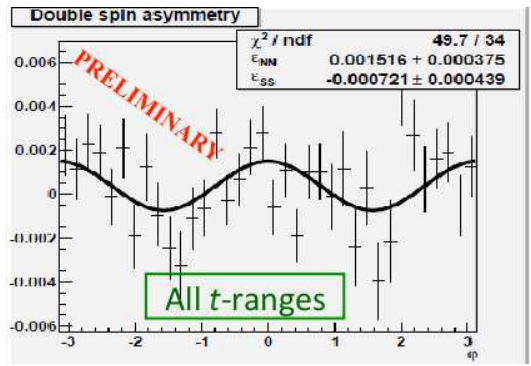}
  \end{center}
  \vspace{-20pt}
  \caption{Preliminary results on double
spin asymmetry. Vertical error bars show statistical uncertainties only} \label{ANN}
  \vspace{-10pt}
\end{wrapfigure}

The preliminary results on double spin raw asymmetries are shown in Fig. \ref{ANN}.
Some effects of the order of 0.001 could be seen. Here we have used
relative luminosities obtained from counts of inelastic triggers produced by the vertex
position detector and beam-beam counters (BBC). However, after more thorough studies,
BBC coincidence counts were proved to be the least sensitive to double spin effects
with negligible statistical uncertainty. The systematic error due to the normalization
uncertainty of BBC coincidence counts on $(A_{NN}+A_{SS})/2$ is at the level of $8.4 \times 10^{-4}$.
\section{Acknowledgments}
 This work was partly supported by the Polish National Science Centre under contract No. UMO-2011/01/M/ST2/04126.

\begin{footnotesize}

\end{footnotesize}
\end{document}